\documentclass{aa}
\usepackage{graphicx}
\usepackage{natbib}
\setcitestyle{aysep={}} 
\usepackage{hyperref}
\usepackage{longtable}
\usepackage{lscape}
\usepackage{color,ulem}
\begin{document}

   \title{Tidal disruption versus planetesimal collisions as possible origins for the dispersing dust cloud around Fomalhaut}

  % \subtitle{Young low-mass binaries}

   \author{Markus Janson\inst{1} \and
          Yanqin Wu\inst{2} \and
          Gianni Cataldi\inst{3,4} \and
          Alexis Brandeker\inst{1}
          }

   \institute{Department of Astronomy, Stockholm University, AlbaNova University Center, 10691 Stockholm, Sweden\\
              \email{markus.janson@astro.su.se}
        \and
           Department of Astronomy and Astrophysics, University of Toronto, 50 St George Street, Toronto, ON M5S 3H4, Canada
         \and
           Department of Astronomy, Graduate School of Science, The University of Tokyo, 7-3-1 Hongo, Bunkyo-ku, Tokyo 113-0033, Japan
         \and
           National Astronomical Observatory of Japan, Osawa 2-21-1, Mitaka, Tokyo 181-8588, Japan
             }

   \date{Received ---; accepted ---}

   \abstract{Recent analysis suggests that the faint optical point source observed around Fomalhaut from 2004--2014  (Fomalhaut b) is gradually fading and expanding, supporting the case that it may be a dispersing dust cloud resulting from the sudden disruption of a planetesimal. These types of disruptions may arise from  catastrophic collisions of planetesimals, which are perturbed from their original orbits in the Fomalhaut dust ring by nearby giant planets. However, disruptions can also occur when the planetesimals pass within the tidal disruption field of the planet(s) that perturbed them in the first place, similar to the Shoemaker-Levy event observed in the Solar System. Given that a gravitationally focusing giant planet has a much larger interaction cross-section than a planetesimal, tidal disruption events can match or outnumber planetesimal collision events in realistic regions of parameter space. Intriguingly, the Fomalhaut dust cloud offers an opportunity to directly distinguish between these scenarios. A tidal disruption scenario leads to a very specific prediction of ephemerides for the planet causing the event. At a most probable mass of 66~$M_\oplus$, a semi-major axis of 117\,AU, and a system age of 400--500\,Myr, this planet would be readily detectable with the \textit{James Webb Space Telescope}. The presence or absence of this planet at the specific, predicted position is therefore a distinctive indicator of whether the dispersing cloud originated from a collision of two planetesimals or from the disruption of a planetesimal in the tidal field of a giant planet.}


\titlerunning{Tidal disruption around Fomalhaut}
\authorrunning{M. Janson et al.}

   \maketitle
%
%________________________________________________________________

\section{Introduction}
\label{s:intro}

The nearby A-type star Fomalhaut has long been known to host a circumstellar debris disk from its strong excess emission at infrared wavelengths \citep{aumann1985}. Spatially resolved images of the disk \citep{holland1998,kalas2005,boley2012} have shown that it is ring-shaped with a large central cavity with a sharp inner edge and a slightly eccentric morphology. These features have been interpreted as evidence for the presence of planets \citep[e.g.][]{quillen2006,chiang2009}; while dust-gas interactions can cause similarly sharp \citep{klahr2005,besla2007} and potentially eccentric \citep[e.g.][]{lyra2013} features without planets, there is not enough gas in the Fomalhaut system for these types of effects to occur there \citep{cataldi2015}.

\textit{Hubble Space Telescope} (HST) images of the Fomalhaut system have revealed a faint optical point source inside of the inner ring edge \citep{kalas2008}, which was initially interpreted as a direct image of a planet, named Fomalhaut b. However, subsequent studies have shown that the interpretation is more complex. The point source lacks an infrared counterpart \citep{janson2012,janson2015}, and it is very blue in the visible \citep{currie2012,galicher2013}. Its orbit also crosses the dust ring, at least in projection \citep{kalas2013}, rather than being nested within it. These characteristics have led to the suggestion that the point source is best explained by a cloud of dust, scattering light from the primary star \citep{janson2012,lawler2015}. A dust cloud of this nature could still be surrounding an unseen planet \citep[e.g.][]{kennedy2011,kenyon2014}. However, the recent discovery that the point source continuously fades and expands over time in HST data from 2004 to 2014 \citep{gaspar2020} lends further support for the dust cloud interpretation and indicates the absence of any central body with a sufficient mass that is holding the cloud together. In this context, the natural explanation for the origin of the dust cloud is a sudden disruption of a planetesimal in the size range of one to a few hundred kilometres \citep[e.g.][]{gaspar2020}.  

One plausible scenario for causing such a disruption is a collision with another, similarly sized, planetesimal \citep{lawler2015,gaspar2020}. However, it is not necessarily a unique scenario. In this paper, we discuss tidal disruption from an interaction with an unseen planet as a possible cause for the dust cloud; and most importantly, we note that Fomalhaut is the first main-sequence system in which the origin for the disruption can be concretely tested observationally.

\section{Origin of the dust cloud}
\label{s:origin}

\subsection{Planetesimal collision (PC)}
\label{s:pc}

Collisions between planetesimals are thought to produce a large fraction of the dust seen in the ring around Fomalhaut and other debris disks \citep[e.g.][]{habing2001}. In this context, this type of collision event is a qualitatively natural interpretation for the transient dust cloud around Fomalhaut. \citet{lawler2015} examine this scenario and find that if there is a large population of dynamically scattered $\sim$100\,km-sized planetesimals in the disk, then cloud-creating events of a sufficient magnitude could occur on decade timescales or even faster, though there would be high uncertainties in the rate. This type of population of planetesimals requires the presence of one or several giant planets inside of the Fomalhaut dust ring, dynamically scattering the planetesimals into higher-eccentricity orbits. 

Here we perform our own estimation of the PC rate. The total dust mass of the dust cloud, in up to 1\,mm grains, is reported to be $\sim 2\times 10^{-8} M_\oplus$ \citep{gaspar2020}. This corresponds to a body with a radius of $30$\,km. In the following, we adopt a total mass for a $100$\,km body to account for the unseen bigger particles, similarly as in \citet{lawler2015}. The two colliding bodies are assigned radii of $R_{\rm target}$ and $R_{\rm bullet}$ with $R_{\rm bullet} < R_{\rm target}$. Given the highly eccentric orbit of the dust cloud, we consider high velocity impacts where an impactor (`bullet')  that is significantly smaller than the target is able to pulverise it.
Moreover, we adopt a power-law size spectrum $dN/dR \propto R^{-q}$, with $q$ typically assumed to be $3.5$ \citep{Dohnanyi(1969)}. Assuming a relative velocity of $\sigma_v$, the smallest bullet that can destroy a target bound by self-gravity is
 \begin{equation}
     \frac{1}{2} \, \frac{4\pi}{3} \rho_{\rm bulk} R_{\rm bullet}^3 \sigma_v^2 \sim \frac{{G (\frac{4\pi}{3} \rho_{\rm bulk} R_{\rm target}^3)^2}}{2 R_{\rm target}}\, 
 \end{equation}
or $R_{\rm bullet}/R_{\rm target} \sim (R_{\rm target}/100\,{\rm km})^{2/3}\,  (0.1\, {\rm km\,s^{-1}}/\sigma_v)^{2/3}$, where we have adopted a bulk density of $\rho_{\rm bulk} \sim 1\, {\rm g\,cm^{-3}}$. In accounting for bodies of similar sizes (same logarithmic decade), we write 
\begin{eqnarray}
    {{N_{\rm bullet}}\over{N_{\rm target}}} & \sim & \left( {{R_{\rm bullet}}\over{R_{\rm target}}}\right)^{1-q} \nonumber \\
   & \sim &
    \left({{R_{\rm target}}\over{100\,{\rm km}}}\right)^{2(1-q)/3} \, 
    \left({{0.1\,{\rm km\,s^{-1}}}\over{\sigma_v}}\right)^{2(1-q)/3}\, .
\end{eqnarray}

For $\sigma_v \sim 1.7$~km/s, which represents the velocity difference between an $e=0.7$ body and a low-eccentricity body at 117\,AU, and $q=3.5$, we find a fairly large number ratio of $\sim 10^2$ for a $100$\,km target. The bullet can be as small as $\sim 15$\,km.

During the time of the hypothesised collision, the dust cloud in the target planetesimal had a highly eccentric orbit \citep{gaspar2020} with $e \sim 0.7$ and $a \sim 330$\,AU. 
So to acquire this orbit, the target, or the centre of mass, had to be excited by the putative planet to at least this value. Such a high eccentricity is not common among all scattered bodies with a range of eccentricities. Typically, planet scattering causes the scattered bodies to random-walk in orbital energies, with each scattering contributing roughly equal energy perturbation (i.e. positive or negative), which leads to a number distribution that is roughly flat in energy, $dn/dE \sim$ constant. 
Since all bodies have to encounter the planet, they all have similar periapses - or apoapses if they orbit inwards of the planet -  $a (1-e) \sim a_p$, where $a_p$ is the planet's semi-major axis. So the number distribution, as a function of the semi-major axis, is as follows:
$    {{dn}\over{da}} = {{dn}\over{dE}} {{dE}\over{da}}\propto {1\over a^2}$ \citep[see, e.g.][]{duncan1987}.
We crudely estimate the high-e portion by integrating over all bodies with $a > 330$\,AU,  $f (a > 330) = n (a > 330)/N_{\rm tot} = {1\over {N_{\rm tot}}} \int dn/da\, da  \sim 0.12 $. 

The collisional probability per orbit can be simply estimated by the total optical depth of colliders, assuming they are spread over a torus of height $z \sim \sigma_i a$, where $a$ is the place of collision and the inclination dispersion $\sigma_i$ can be taken to be of $\sim 0.1$ to account for the heated belt. The rate of collisions that can produce the observed dust cloud is then
\begin{eqnarray}
\Gamma_{\rm PC} & \sim &
{2\over{P_{\rm orb}}}\, f(a > 330\,{\rm AU}) \times N_{\rm target} \times {{N_{\rm bullet} \pi R_{\rm target}^2}\over{\sigma_i 2 \pi a^2}} \nonumber \\
& \sim & {0.04/{\rm decade}}\, \times 
\left({{N_{\rm target}}\over{10^8}}\right)^2 \, 
\left({{R_{\rm target}}\over{100\,{\rm km}}}\right)^{1/3}
\nonumber \\
& & \times \left({{\sigma_v}\over{1.7\,{\rm km\,s^{-1}}}}\right)^{5/3}\, \left({{P_{\rm orb}}\over{4400\,{\rm yr}}}\right)^{-1} \nonumber \\ 
& & \times \left({{f(a> 330\,{\rm AU})}\over{0.12}}\right) \, 
\left({{\sigma_i}\over{0.1}}\right)^{-1} \, ,
\label{eq:gammacol}
\end{eqnarray}
where $q=3.5$. The relevant orbital period is that of the highly-eccentric body, $P_{\rm orb} \geq 4400$\,yr, and $N_{\rm target}$ is the total number of bodies with $R \sim R_{\rm target}$ (within a decade). We have chosen a target number, $10^8$, that is consistent with the \citet{lawler2015} model. This estimated rate is very sensitive to the value of the size-index $q$. Taking $q \sim 2$, the observed size index for Kuiper belt bodies below $100$\,km in radius \citep[e.g.][]{bernstein2004,fraser2009}, the rate drops by more than an order of magnitude to $\Gamma_{\rm PC} \sim 0.002$ events/decade, when all other parameters are the same as those in 
eq.~\ref{eq:gammacol}. Our rate falls by about an order of magnitude below the one in \citet{lawler2015}, primarily because we restrict the colliding bodies so they have a high eccentricity, as is observed for the dust cloud.

While a $\sim$4\% event occurring during the approximate decade across which Fomalhaut has been monitored requires a bit of luck, it is not an unreasonable incidence, so we consider PC to be a potentially feasible mechanism to generate the dust cloud. While we discuss a particularly interesting test of this scenario in Sect. \ref{s:obs}, the collisional model has a number of consequences, some of which we describe below:

\begin{enumerate}

    \item Total belt mass. The total mass of the planetesimal belt, which is composed of bodies up to a size of $R_{\rm max}$, is
\begin{equation}
    M_{\rm belt} = \int^{R_{\rm max}}_{R_{\rm min}} {{4\pi}\over 3} \rho_{\rm bulk} R^3 {{dn}\over{dR}} \, dR  \, 
%\sim n_{R_{\rm max}} {{4 \pi}\over 3} \rho_{\rm bulk} R_{\rm max}^3 
\sim 70 M_\oplus  \left({{R_{\rm max}}\over{100\,{\rm km}}}\right)^{4-q}\,     
\end{equation}
or $10^3$--$10^4$ times higher than the mass of the current Kuiper belt.  This applies for a size distribution that is top-heavy ($q < 4$). The total solid mass required in the planetesimal belt is thus remarkably high, especially since it is likely that $R_{\rm max} \gg 100$\,km.

\item Dimmer events. Eq. \ref{eq:gammacol} implies that smaller bodies should be destroyed at a much higher rate, as $\Gamma_{\rm PC} \propto R_{\rm target}^{2+8(1-q)/3}$, while the brightness of such events are dimmer by a factor of $R_{\rm target}^3$. For $q=3.5$, if we witness one $100$\,km class destruction over the past decade, events that are dimmer by a factor of $10$ should be more populous by a factor of $36$. Unfortunately, since the $S/N$ of HST dust cloud measurements has typically been $\sim$10 or lower, such events would not be easily detectable. However, with a greater image depth and contrast, they could be a common occurrence. 

\item Post-disruption dispersal. Assuming that the target is barely unbound by the small impactor, the outflow should be of the order of the escape velocity from the large body \citep[see, e.g.][]{lawler2015}.\footnote{As opposed to our interpretation here, \citet{gaspar2020} infer that the dispersal velocity is the collisional velocity. If so, it will require both colliding bodies to move at low relative velocities, i.e. both on highly eccentric orbits, which is a much less likely scenario. Moreover, it requires an impactor that is comparable in size to the target, which is a very infrequent occurrence.} \citet{gaspar2020} report a dispersal velocity of $236$\,m\,s$^{-1}$ by studying the size evolution of the dust cloud, but a larger value ($616$\,m\,s$^{-1}$) from photometric modelling. The former is comparable, but somewhat larger, than the escape velocity of a $100$\,km body.

\end{enumerate}

\subsection{Tidal disruption (TD)}
\label{s:td}

If a planetesimal passes too close to a massive planet, it can be torn apart by tidal forcing. Indeed, the aftermath of such an event has been observed in our own Solar System, via the fragmentation of Shoemaker-Levy 9 \citep{shoemaker1993} in the gravitational field of Jupiter. Tidal disruption of asteroidal objects is also thought to be the mechanism responsible for the debris disks seen around white dwarfs \citep[e.g.][]{farihi2009,barber2012}. In that context, the disruption is caused by the white dwarf itself; see \citet{veras2016} for a theoretical review.

Since a massive planet must be present anyways within dynamical reach of the Fomalhaut planetesimal belt in order to excite a scattered population required for the PC scenario to work, it is relevant to assess not only at which rate the planetesimals collide with each other, but also at which rate they undergo tidal disruptions by an interaction with the planet. On the one hand, PC is favoured since each planetesimal could collide with a large number of other planetesimals, while there is only one or a few planets with which it could undergo a TD interaction. On the other hand, the cross-section of TD interactions with the planet is much higher than the cross-section of planetesimal collision, for two reasons: (1) A giant planet is strongly gravitationally focusing, while a planetesimal is not; and (2) the tidal destruction radius of the planet is a couple of times larger than its physical radius, which is relevant regarding collision, and significantly larger than the collisional radius of a planetesimal. In the following, we also discuss the expected morphology and orbit of this type of event.

Here, we follow the discussion in \citet{cataldi2018} to estimate the TD rate in Fomalhaut. We take the semi-major axis of the unseen planet that is hypothesised to be responsible for the TD event as $a_p = 117$\,AU, its mass as $M_p = 66\,M_\oplus$ (see Sect. \ref{s:obs}), and a tidal disruption radius of $2 R_{\rm Neptune} \sim 5\times 10^4$\,km. Accounting for gravitational focusing, $N$ planetesimals on crossing-orbits with the planet can be tidally destroyed at a rate of 
\begin{eqnarray}
\Gamma_{\rm TD} & \sim &
{2\over{P_{\rm orb}}}\,  N
\times {{\pi R_{\rm TD}^2 \left(v_{\rm esc} / \sigma_v\right)^2}\over{\sigma_i 2 \pi a^2}} \nonumber \\
& \sim & {0.04/{\rm decade}}\, \times 
\left({{N}\over{10^8}}\right) \, 
\left({{R_{\rm TD}}\over{5\times 10^4\,{\rm km}}}\right)\, 
\left({{M_p}\over{66 M_\oplus}}\right) \nonumber \\
& & \times \left({{P_{\rm orb}}\over{1200\,{\rm yr}}}\right)^{-1} \, 
\left({{\sigma_i}\over{0.1}}\right)^{-3} \, ,
\label{eq:gammatd}
\end{eqnarray}
where the escape velocity at the tidal radius $R_{\rm TD}$ is $v_{\rm esc}= \sqrt{2 G M_p/R_{\rm TD}}$, and we have assumed a velocity dispersion for the orbit-crossing bodies to be $\sigma_v \sim \sigma_i \times v_{\rm kep} \sim 1.4$\,km\,s$^{-1}$, which is appropriate for the scattered planetesimal population in the disk. This is much smaller than $v_{\rm esc} \sim 23$\,km\,s$^{-1}$, allowing for strong gravitational focussing. Our estimated rate is not sensitive to the power-law index $q$, unlike the rate for collisional destruction.

% Rates
\begin{figure}[htb]
\centering
\includegraphics[width=8.5cm]{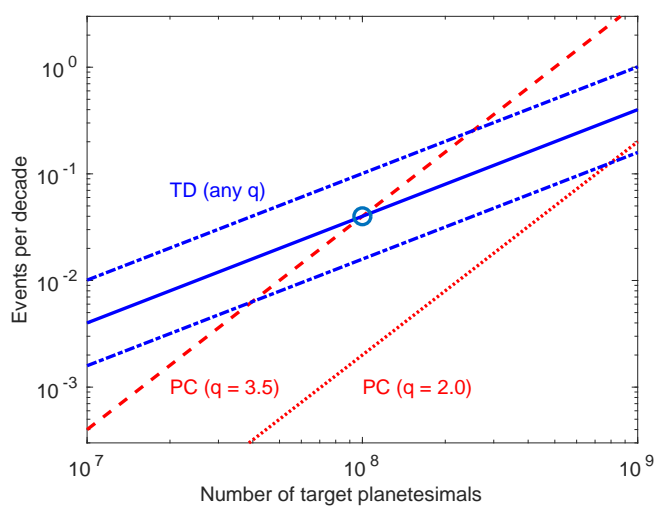}
\caption{Calculated rates for PC and TD as a function of the two most uncertain parameters: the number of possible target bodies for the disruption ($N_{\rm target}$ for PC, $N$ for TD) and the power-law index $q$. Blue solid line: TD rate, which is insensitive to $q$. Blue dash-dotted lines: The TD rate, if the planetary mass is shifted up or down by a factor 2, illustrates the rate dependence on the mass of the disrupting body. Red lines: PC rate, for $q = 2$ (dotted line) and $q = 3.5$ (dashed line). The circle marks $10^8$ target bodies, which is of the same order as was estimated in \citet{lawler2015}.}
\label{f:rates}
\end{figure}

We make a few remarks concerning this type of disruption event beyond what is discussed in Sect. \ref{s:obs}.

\begin{enumerate}

\item No expectation of dust around the planet: in our case, $\sigma_v \gg v_{\rm esc}'$, surface escape velocity of the planetesimal, ($v_{\rm esc}' \sim 0.1$\,km\,s$^{-1}$ for a 100\,km body). This suggests that only a small fraction of the original orbital energy is used to tear the body apart and most of the debris still leave the planet on the original hyperbolic orbit, relative to the planet. This scenario is markedly different from the commonly discussed stellar tidal disruption by a massive black hole \citep[e.g.][]{rees1988}, where $\sigma_v \ll v_{\rm esc}'$ and where roughly half of the stellar mass becomes bound to the black hole. Likewise, it is different from the case of tidal disruption around white dwarfs, where the tidally disrupted asteroidal body is in a bound orbit around the white dwarf, as opposed to the parabolic and hyperbolic orbits in the other scenarios. In our application, the offending planet is not expected to be surrounded by a large dust cloud.

\item Post-disruption morphology: \citet{richardson1998} simulated the tidal disruption of asteroids by Earth. Our event, with a slow approach and a large impact parameter, likely belongs to a category they called M-class (M for mild). The end product does not resemble a `string-of-pearls', as the famed Comet Shoemaker-Levy 9), but more a `pinwheel' where the disrupted body is spun up to possibly a break-up speed and starts to shed material nearly isotropically. If so, the mass outflow would have a velocity of the order of the surface escape velocity, which would explain the inferred dispersal velocity for the dust cloud. Simulations of disruption around Saturn, which is still a closer analogue to our application, give consistent results \citep{leinhardt2012}.

\item High eccentricity: The close approach that leads to the disruption can help to explain the dust cloud's highly eccentric motion. Unlike the PC case, the parent planetesimal does not need to be a member of the extreme-eccentricity population because the very approach that destroys it also provides a gravitational sling-shot, typically doubling its eccentricity.

\end{enumerate}

In summary, tidal disruption of a 100\,km body may give rise to the correct morphology and orbital characteristics for Fomalhaut b. The disrupted body leaves the planet largely as a collected unit and then slowly disperses over time. Under the same assumptions for the underlying system properties, our calculated rate of $\Gamma_{\rm TD} \sim$0.04 event per decade coincides with that for PC events if $q = 3.5$, whereas it is significantly larger than for PC events if $q = 2$, see Fig. \ref{f:rates}. The uncertainties are still on the level of several orders of magnitude; however, they depend on a fair number of unknowns. Additional observational constraints are therefore necessary to distinguish between these scenarios as origins for the Fomalhaut dust cloud.

\section{Distinguishing the scenarios}
\label{s:obs}

A particularly appealing aspect of examining TD versus PC in the Fomalhaut system is that it will be concretely testable in the near future with the \textit{James Webb Space Telescope} (JWST). Indeed, Fomalhaut will be observed with coronagraphy in JWST GTO programme 1193 (PI: C. Beichman, hereafter JWST-1193), with both NIRCAM at $\sim$4.4\,$\mu$m and MIRI at $\sim$15\,$\mu$m. Here we use that programme as a template when examining observational distinctions between the two scenarios. The work of \citet{gaspar2020} allows one to backtrace the dust cloud to its origin at $\Delta x = 8.583^{\prime \prime}$ (west) and $\Delta y = 9.143^{\prime \prime}$ (north), at a most likely date of 2004.03. If TD is responsible for the dust cloud, then a planet must have been present at this time and place for the event to have occurred. If this planet is also responsible for shaping the inner gap of the disk, which we assume for now, then it shares the same apsides as the disk and its separation to the disk edge implies a mass of $\sim$66~$M_\oplus$ from the interpolation of the results in \citet{chiang2009}. We can thus make predictions for both the position and brightness of the planet at any given epoch. This is discussed in more detail in the following sections, where we consistently adopt a distance to Fomalhaut of 7.7\,pc \citep{vanleeuwen2007} and an age and stellar mass of $\sim$400--500\,Myr and 1.92\,$M_{\rm sun}$, respectively \citep{mamajek2012}.

% Ephemeris
\begin{figure}[htb]
\centering
\includegraphics[width=8.5cm]{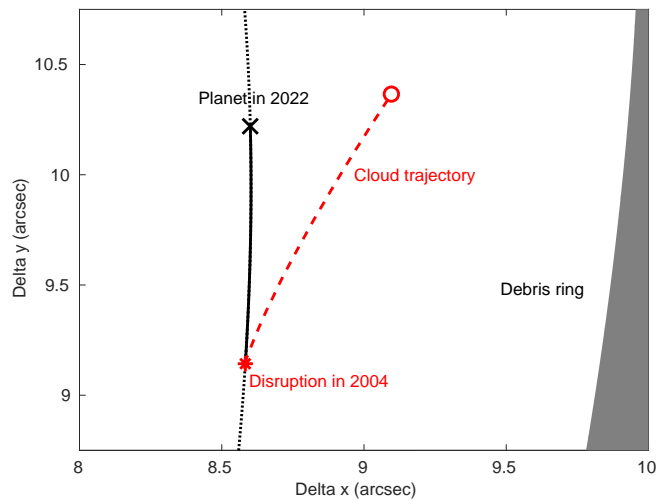}
\caption{Prediction for the location of the hypothesised TD-causing planet in October 2022 (`x'-mark) in the best-fit case, along with its orbit (solid line) since the disruption in 2004 (asterisk). The best-fit trajectory of the dust cloud from \citet{gaspar2020} is also shown as a dashed line. North is up, east is to the left.}
\label{f:ephemeris}
\end{figure}

\subsection{Ephemeris for the unseen planet (UP)}
\label{s:ephemeris}

The Fomalhaut disk parameters were determined in \citet{kalas2013} based on the following bisector analysis: semi-major axis $a_{\rm ring}$ of 141.77\,AU, inclination $i = -66.1\degr$, ascending (or descending) node $\Omega = 156.1\degr$, eccentricity $e = 0.1$, and argument of periastron $\omega = 29.6\degr$. The inner edge of the disk is at $a_{\rm inner} = 136.28$\,AU. By assuming the same orbital parameters for the apsidally locked planet inferred in the TD hypothesis, but scaling the semi-major axis to match the origin of the dust cloud, we find $a_{\rm UP} = 117.58$\,AU and consequently an orbital period of $P_{\rm UP} = 1218.2$\,yr. With these orbital parameters, we determine the time of periastron that enables UP to match up in time with the cloud origin: $t_{\rm p} = 2431248$\,d (JD). This assumes a counter-clockwise motion, which we deem most probable based on the direction of the dust cloud. The listed parameters can now be used to predict the location of UP at any given point in time. The most relevant date to evaluate an ephemeris prediction is the date when Fomalhaut will be observed in JWST-1193. We have evaluated this by examining the publicly available observing set-up for the programme. The exact date is obviously unknown, not only because the launch date of JWST is uncertain, but due to the intricate combination of constraints for the observations; they can only be executed around October 20 each year, within a few days of margin. We choose 2022 October 20 as the first occasion at which Fomalhaut could plausibly be observed, which gives ephemerides of $\Delta x = 8.600^{\prime \prime}$ (west) and $\Delta y = 10.220^{\prime \prime}$ (north), see Fig.~\ref{f:ephemeris}. If observed in nearby subsequent years on the same date, the ephemerides change by approximately 55\,mas northward and 1\,mas eastward per year. The uncertainty in the prediction is dominated by the uncertainty in the time of the cloud origin based on the cloud expansion, which is $\pm$0.91\,yr \citep{gaspar2020}, corresponding to $\sim$100\,mas in uncertainty in the ephemerides. If the putative planet is observed within the near future, then its conspicuous location close to the disruption zone is, by itself, evidence for it being the tidal disruptor. However, even if the planet is detected after a longer timescale, it can still be efficiently traced back to the dust cloud origin, as long as two epochs are acquired. For example, if two astrometric data points are acquired with a 1 mas precision (each), over a baseline of one year, then the orbital velocity can be determined with $\sim$2\% precision and the origin can be traced to within $\sim$100 mas even after $\sim$70 years.

\subsection{JWST observability}
\label{s:jwst}

\begin{figure}[htb]
\centering
\includegraphics[width=8.5cm]{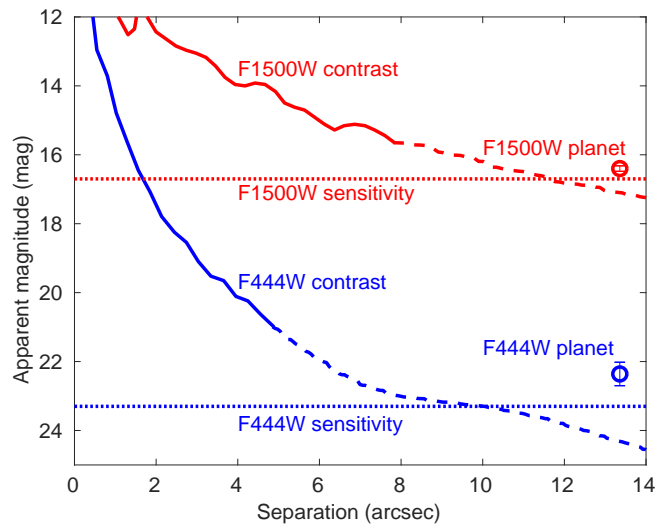}
\caption{Detection limits for JWST-1193 with NIRCAM (blue lines) and MIRI (red lines). The expected brightness for the hypothesised 66~$M_\oplus$ planet is also shown for each filter as circles. The dashed lines are the extrapolated portions of the contrast curves (see text).}
\label{f:contrast}
\end{figure}

We evaluate the detectability by considering the contrast and sensitivity limits separately and also by requiring that any planet brightness must fulfil both in order to be detectable. We consistently use 5$\sigma$ as a criterion; although with a pre-determined position, it may well be sufficient with a lower detection threshold in reality. For sensitivity limits we use the JWST Exposure Time Calculator (ETC) with the observational settings of JWST-1193. We do not use coronagraphic versions of the ETC because they cannot handle the large separation of the Fomalhaut system; instead, we calculate sensitivity to isolated objects with the F444W filter for NIRCAM and the F1500W filter for MIRI, manually applying the 62\% throughput of the MIRI coronagraph as a correction to the results. The NIRCAM observations include two different readout settings; here we consider a sum of the two, assuming that the $S/N$ adds in quadrature. As a basis for the contrast evaluation, we use simulated curves from \citet{beichman2010} for NIRCAM and \citet{boccaletti2015} for MIRI; however, since they both cut off at separations smaller than our region of interest, we extrapolate the curves by simulating the PSF in each relevant band with the WebbPSF tool and by assuming that the achievable contrast remains fixed with respect to the local PSF level at large separations. The results are shown in Fig. \ref{f:contrast}, where we also show the predictions for UP, the hypothesised TD-causing planet. These were evaluated using the mass-luminosity models from \citet{linder2019} for a 66\,$M_\oplus$ planet at the separation calculated in Sect. \ref{s:ephemeris}. At 22.0--22.7\,mag in F444W and 16.3--16.5\,mag in F1500W, the planet would be easily detectable in both filters. Even lower-mass planets would also be detectable. Since the model grids become incomplete at 400--500\,Myr below 50\,$M_\oplus$, the exact detectable mass cannot be determined, but it is certainly below 50~$M_\oplus$. We also note that the JWST-1193 observations are not particularly deep, and the observations are sensitivity-limited at the relevant separation, so in the future it would be possible to go even lower in mass with a deeper JWST run. 

% Contrast curves

\section{Discussion and conclusions}
\label{s:discussion}

The TD scenario makes a clear prediction on the position of the planet when the dust cloud was disrupted, which has an uncertainty of $\sim
100$ mas \citep[][]{gaspar2020}. If we adopt some straightforward assumptions on the planet orbit, this yields the position of the planet at the JWST epoch. This makes the JWST observations a compelling test. Here, we discuss our assumptions and some implications.

For the ephemeris calculation in Sect. \ref{s:ephemeris}, we have assumed an orbit and a mass for the planet that are consistent with sculpting the eccentric debris ring. If we relax these values, it may impact our predictions by varying degrees. First, we have assumed a counter-clock-wise (CCW) orbit based on the cloud motion. In principle, a clock-wise (CW) orbit is also possible. In this circumstance, the orbital elements remain the same apart from, for example, a flip of $i$ by 180 deg (to 113.9\degr) and a new $t_{\rm p} = 2313672$\,d, giving an ephemeris of $\Delta x = 9.169^{\prime \prime}$ (west) and $\Delta y = 8.777^{\prime \prime}$ (north) on 2022 October 20. For subsequent years, it would move by approximately 56\,mas south and 3\,mas west per year. The slightly smaller separation than the one in the CCW case (12.7$^{\prime \prime}$ for CW versus 13.4$^{\prime \prime}$ for CCW) does not have any significant impact on the observability. Second, it is, in principle, possible for a non-apsidally locked planet to cross the origin point and cause tidal disruption. This, however, would then require the presence of another planet to explain the debris ring morphology. Lastly, given the range of periapses of the scattered planetesimal population, planets at a smaller semi-major axes can also be responsible for TD events. In particular, if this type of planet is more massive, it would enhance the rate of tidal disruption (eq.~\ref{eq:gammatd}). A higher mass would also benefit detectability. Conversely, if no planet is found or if the planet orbit is inconsistent with tidal disruption, this would rule out the TD scenario and the case for PC is significantly strengthened.

Regardless of whether TD is the actual scenario behind this particular event, it is a mechanism that requires consideration in a wide range of debris disk-related applications. The number of available planetesimals $N$ is a key parameter in this context. Since PC scales as $N^2$, while TD as $N$, and since $N$ rises steeply with a decreasing body size, we expect PC to be more dominant for events involving small bodies and TD to be more dominant for larger bodies. The applications of the latter potentially include our massive dust cloud in Fomalhaut and large asymmetries in debris disks \citep{cataldi2018}.

In either scenario, a massive planet that is capable of scattering a massive planetesimal belt is required. It is even possible that not one, but multiple planets are involved in the process, as is the case in our Solar System. Therefore, both the PC and TD scenarios predict massive and potentially JWST-detectable planets within the gap. It is the ephemerides of those planets that can provide the distinction for the origins of the dust cloud.

In summary, JWST can provide an observational distinction between planetesimal collisions versus tidal disruption as the underlying cause for the expanding dust cloud around Fomalhaut. The predicted ephemeris is based on the observed properties of the cloud and is independent of the large uncertainties in the rate calculations discussed in Sections \ref{s:pc} and \ref{s:td}. It is therefore more straightforward and robust than assessing the relative probabilities based on these types of calculations.

\begin{acknowledgements}
M.J. gratefully acknowledges funding from the Knut and Alice Wallenberg Foundation. We made use of CDS and NASA/ADS services for the purpose of this study.
\end{acknowledgements}

\end{document}